\documentclass{elsart3}
\usepackage{graphicx}
\usepackage{amssymb}

\begin{document}

\begin{frontmatter}

\title{Strong- and weak-coupling mechanisms for pseudogap in electron-doped cuprates}

\author[uds,tstu]{V.~Hankevych,}
\author[uds]{B.~Kyung,}
\author[l2mp]{A.-M.~Dar\'e,} 
\author[uds]{D.~S\'en\'echal,} and 
\author[uds]{A.-M.~S.~Tremblay}
\address[uds]{D\'{e}partement de physique and Regroupement qu\'{e}b\'{e}cois sur les
mat\'{e}riaux de pointe,\\ Universit\'{e} de Sherbrooke, Sherbrooke,
Qu\'{e}bec J1K 2R1, Canada}
\address[tstu]{Department of Physics, Ternopil State Technical University, 56 Rus'ka
St., UA-46001 Ternopil, Ukraine}
\address[l2mp]{L2MP, 49 rue Joliot Curie BP 146, Universit\'{e} de Provence, 13384
Marseille, Cedex 13, France}

\begin{abstract}
Using the two-particle self-consistent approach and cluster perturbation
theory for the two-dimensional $t$-$t'$-$t''$-$U$ Hubbard model, we
discuss weak- and strong-coupling mechanisms for the pseudogap observed
in recent angle resolved photoemission spectroscopy on electron-doped
cuprates. In the case of the strong-coupling mechanism, which is more relevant
near half-filling, the pseudogap can be mainly driven by short range correlations
near the Mott insulator. In the vicinity of optimal doping,
where weak-coupling physics is more relevant, large antiferromagnetic
correlation lengths, seen in neutron measurements, are the origin of
the pseudogap. The $t-J$ model is not applicable in the latter case.
\end{abstract}

\begin{keyword}
Pseudogap \sep the Hubbard model \sep electron-doped cuprates \sep
antiferromagnetic fluctuations

\PACS 
74.72.-h \sep 71.10.Fd \sep 71.27.+a
\end{keyword}
\end{frontmatter}

Angle resolved photoemission spectroscopy~\cite{Damascelli2003} (ARPES)
provides deep insight into the nature of high-temperature superconductors.
In particular, it has revealed the failure of Fermi liquid theory to
describe single-particle excitations in these systems. Contrary to the
quasiparticle concept of Fermi liquid theory, certain segments of the
would-be Fermi surface are almost gapped. This is the so-called pseudogap
phenomenon. In particular, recent ARPES measurements~\cite{Armitage2001} on
Nd$_{2-x}$Ce$_{x}$CuO$_{4}$ have shown that, in contrast to the hole-doped
cuprates, lightly electron-doped (e-d) ones have a large spectral weight
near ($\pi ,0$). With further doping towards optimal doping, spectral weight
also appears around the zone diagonals, leaving hot spots (regions with
large scattering) in between ($\pi ,0$) and ($\pi /2,\pi /2$) where the
non-interacting Fermi surface intersects the antiferromagnetic (AFM) zone
boundary. Theoretical explanation of these experimental data is still an
open question.

In the present paper we discuss two approaches for the pseudogap in e-d
cuprates: a strong and a weak-coupling one. In the case of the
strong-coupling mechanism~\cite{st2004}, which is more relevant near
half-filling, the pseudogap is mainly driven by short-range correlations
near the Mott insulator. In the vicinity of optimal doping, where a
weak-coupling mechanism is more appropriate, large AFM correlation lengths
are the origin of the pseudogap~\cite{khdt2004}.

We use two different methods, the two-particle self-consistent (TPSC)
approach~\cite{vt1997,allen03} and cluster perturbation theory~\cite{spp2000}
(CPT), for the single-band Hubbard model on a square lattice with a
repulsive local interaction $U$ and nearest $t$, next-nearest $t^{\prime }$
and third-nearest $t^{\prime \prime }$ neighbour hoppings. The former method
is based on a self-consistent determination of the irreducible vertices that
enter dynamical susceptibilities (spin-spin and density-density). This is
done by enforcing the Pauli principle, conservation laws for spin and charge
fluctuations, and important sum rules. These results are then used to obtain
an improved approximation for the single-particle self-energy. TPSC has been
extensively checked against Quantum Monte Carlo simulations~\cite%
{vt1997,allen03,malkpvt00}. The CPT approach is based on exact
diagonalization of finite clusters that are coupled through strong-coupling
perturbation theory. The CPT results were calculated on $4\times 4$
clusters. Since TPSC is valid in the weak to intermediate coupling regime,
while CPT is more reliable for intermediate to strong coupling, we can study
all values of $U$ to gain insight into two mechanisms for the pseudogap in
e-d cuprates.

\textbf{Weak-coupling pseudogap:} In the vicinity of optimal doping, the
absence of zero energy excitations around ($\pi /2,\pi /2$) is a general
result of strong-coupling calculations~\cite{st2004,toh2001} for $t$-$%
t^{\prime }$-$t^{\prime \prime }$-$U$ and $t$-$t^{\prime }$-$t^{\prime
\prime }$-$J$ models. This is consistent even with weak-coupling
calculations~\cite{klt2003,kr2003} for the Hubbard model when the coupling
is increased towards the value of the bandwidth. On the other hand, several
papers~\cite{st2004,khdt2004,klt2003,Kusko2002,Markiewicz2003} have shown
that the spectral weight near ($\pi /2,\pi /2$), as seen in experiment~\cite%
{Armitage2001}, shows up at optimal doping when the Hubbard coupling $U$ is
somewhat smaller than the bandwidth. Therefore, one concludes that
strong-coupling physics is not relevant for the pseudogap in e-d cuprates
near optimal doping, and that a weak to intermediate coupling mechanism is
appropriate. In other words, the $t$-$J$ model does not describe the physics
of e-d cuprates near optimal doping. Further evidence for this is the
disappearance at $x=0.15$ of the lower Hubbard band that is present at $%
x=0.04$ in ARPES data~\cite{Armitage2001} on Nd$_{2-x}$Ce$_{x}$CuO$_{4}$.
One can derive the $t$-$J$ model from the Hubbard model only when both upper
and lower Hubbard bands are well defined.

At weak coupling, large AFM correlation lengths, seen in neutron
measurements~\cite{Mang2003}, are the driving force for the pseudogap~\cite%
{khdt2004}. The physical mechanism is that electrons on a planar lattice
suffer scattering by AFM fluctuations which have large phase space in two
dimensions \cite{vt1997,v97}. Thus, those quasiparticles in regions of the
Fermi surface that can be connected by the AFM vector (so called hot spots)
do not survive scattering by these strong AFM fluctuations. This theory~\cite%
{khdt2004} explains in detail the ARPES results mentioned above, as well as
the temperature dependent correlation length measured by neutron scattering~%
\cite{Mang2003}. Ref.~\cite{khdt2004} also makes a few predictions for
ongoing experiments: (a) The ARPES pseudogap found at low temperatures
should be seen even in the paramagnetic phase up to the temperature when the
AFM correlation length becomes smaller than the single-particle thermal de
Broglie wavelength $\xi _{th}=\hslash v_{F}/\pi k_{B}T$. The corresponding
pseudogap temperature $T^{\ast }$ is close to that found in optical
experiments~\cite{Onose2001}. (b) For $T<T^{\ast }$ and for $T$ somewhat
larger than $T^{\ast }$, the characteristic spin fluctuation energy in
neutron scattering experiments is smaller than the thermal energy
(renormalized classical regime) and the spin fluctuations are overdamped
near $T^{\ast }$.

In the spirit of Ref.~\cite{khdt2004} we also compare the renormalized Fermi
velocities at optimal doping $n=1.15$ along the $(\pi ,0)$-$(\pi ,\pi )$
direction and along the zone diagonal with the corresponding ARPES data~\cite%
{armitage2003} on Nd$_{1.85}$Ce$_{0.15}$CuO$_{4}$. Using the experimental
renormalization factors and bare Fermi velocities~\cite{armitage2003}, the
experimental renormalized Fermi velocities are $3.31\times 10^{5}$~m/s and $%
3.09\times 10^{5}$~m/s along the zone diagonal and along the $(\pi ,0)$-$%
(\pi ,\pi )$ direction, respectively. The corresponding renormalized Fermi
velocities obtained by TPSC are $3.27\times 10^{5}$~m/s and $2.49\times
10^{5}$~m/s, respectively. The agreement is very good, particularly along
the diagonal direction. The bare Fermi velocities are renormalized in TPSC
by roughly a factor of two.

\begin{figure}[tbp]
\includegraphics[width=7.0cm]{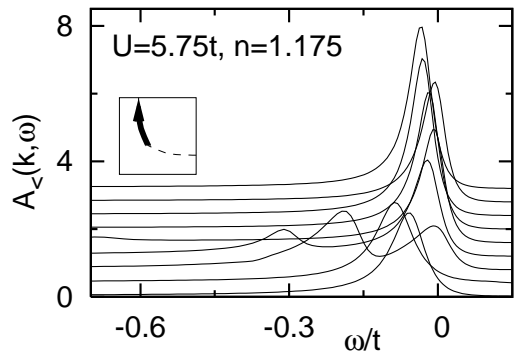}
\caption{Energy distribution curves $A_{<}(\mathbf{k},\protect\omega )\equiv
A(\mathbf{k},\protect\omega )f(\protect\omega )$ along the Fermi surface
shown in the inset for $n=1.175$, $U=5.75t$. The lines are shifted by a
constant for clarity. Band parameters are $t^{\prime }=-0.175t$, $t^{\prime
\prime }=0.05t$ while $T=t/50$.}
\label{spectral_function}
\end{figure}

\begin{figure}[tbp]
\includegraphics[width=7.5cm]{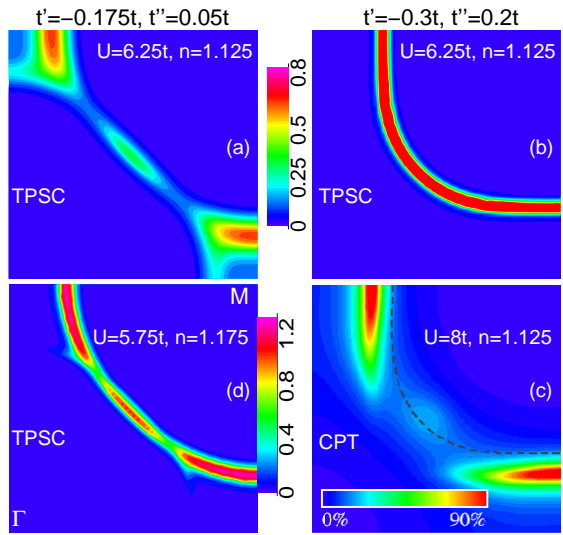}
\caption{(color) Fermi surface plots $A_{<}(\mathbf{k},0)$ in the first
quadrant of the Brillouin zone by two methods: TPSC and CPT, $\Gamma
=(0,0),\ \mathrm{M}=(\protect\pi ,\protect\pi )$. The TPSC plot in part (a)
and (b) are at $T=0.05t,$ that in part (d) is at $T=0.02t$ while the CPT
plot in part (c) is at $T=0.$}
\label{FS_plot}
\end{figure}

In Fig.~\ref{spectral_function} we present the single-particle spectral
weight obtained by TPSC for filling $n=1.175$ and temperature $T=t/50$ for
the same values of $U$ and band parameters as in Ref.~\cite{khdt2004}. Here, 
$A(\mathbf{k},\omega )$ is multiplied by the Fermi-Dirac distribution
function $f(\omega )$. One can see that $A_{<}(\mathbf{k}_{F},\omega )$ is
peaked at zero energy near ($\pi ,0$) and ($\pi /2,\pi /2$), and is shifted
away from the Fermi energy (pseudogaped) towards higher binding energies at
hot spots where the Fermi surface intersects the AFM Brillouin zone
boundary. The calculated Fermi surface plot $A_{<}(\mathbf{k},0)$ that
corresponds to this case (Fig.~\ref{FS_plot}(d)) shows, at hot spots, the
zero energy suppression of spectral weight by AFM fluctuations. Thus, we
predict that the pseudogap induced by AFM fluctuations should be
experimentally seen up to $18\%$ electron doping, and should disappear for
larger dopings~\cite{khdt2004}. Note however that $18\%$ doping in our
calculations may be equivalent to $15\%$ doping in reduced samples if we
take the point of view of Ref.~\cite{Mang2003} that the role of reduction
can be modeled as a $\Delta x\approx 0.03$ shift with respect to nominal Ce
concentration. Then our data for the Fermi surface plot and spectral
function at $n=1.175$ should be compared with the ARPES ones for reduced
samples near $x=0.15$ Ce doping. The semi-quantitative agreement of Ref.~%
\cite{khdt2004} is preserved since Fermi surface plots and energy
distribution curves for $n=1.15$ (see Figs.~1,2 of Ref.~\cite{khdt2004}) and 
$n=1.175$ (shown here) look similar. However, in better agreement with
experiment, the theoretical pseudogap feature for $15\%$ doping is slightly
more pronounced and extends over a broader region in $\mathbf{k}$ space than
that for $17.5\%$ doping. For comparison of our correlation lengths with
neutron measurements data on $15\%$ doped as grown samples we should use $%
n=1.15$ as was done in Ref.~\cite{khdt2004}. Note however that Fig.3 of Ref.~%
\cite{Onose2001} may lead us to argue that doping corresponds to nominal Ce
concentration only in reduced samples. The role of reduction and Ce alloying
on doping remains to be clarified.

As shown in the Fermi surface plot obtained by TPSC [Fig.~\ref{FS_plot}(a)],
for $t^{\prime }=-0.175t,\ t^{\prime \prime }=0.05t$, strong AFM
fluctuations at $12.5\%$ doping cause the suppression of spectral weight not
only at hot spots, but also along a large segment of the Fermi surface near $%
(\pi /2,\pi /2)$. This is consistent with strong-coupling calculations~\cite%
{st2004}. The AFM correlation length is about 40 lattice spacings for this
plot, which is larger than $\xi _{th}$, and the spin susceptibility at $(\pi
,\pi )$ is much larger than the noninteracting one. By contrast, for large
values of $|t^{\prime }|=0.3t$ and $t^{\prime \prime }=0.2t$, one can see in
Fig.~\ref{FS_plot}(b) the Fermi liquid like (uniform) distribution of
spectral weight in $\mathbf{k}$ space that results from the strong
suppression of AFM fluctuations by frustration. Here the AFM correlation
length is about two lattice spacings, and the spin susceptibility at an
incommensurate wave vector $(\pi -\delta ,\pi )$ is just a few times larger
than the noninteracting one. This again shows that large AFM correlation
lengths are the driving force for the pseudogap in the weak-coupling case.

\textbf{Strong-coupling pseudogap:} An additional type of physics becomes
relevant in the strong-coupling regime, which we argued is more appropriate
when electron doping is decreased towards half-filling. In this limit, short
range correlations are sufficient to create a pseudogap even in the absence
of long AFM correlation lengths~\cite{st2004}. As shown in the Fermi surface
plot obtained by CPT [Fig.~\ref{FS_plot}(c)], for large values of $%
|t^{\prime }|=0.3t$ and $t^{\prime \prime }=0.2t$ (enough to frustrate
antiferromagnetism in TPSC) the pseudogap does occur at $12.5\%$ doping
around the nodal direction as a result of strong coupling, whereas it is
absent at weak coupling for the same values of $t^{\prime }$ and $t^{\prime
\prime }$ [see Fig.~\ref{FS_plot}(b)]. We cannot rule out the existence of
long AFM correlation lengths at strong coupling with CPT because the finite
cluster size precludes such correlations. 

\begin{figure}[tbp]
\includegraphics[width=7.5cm]{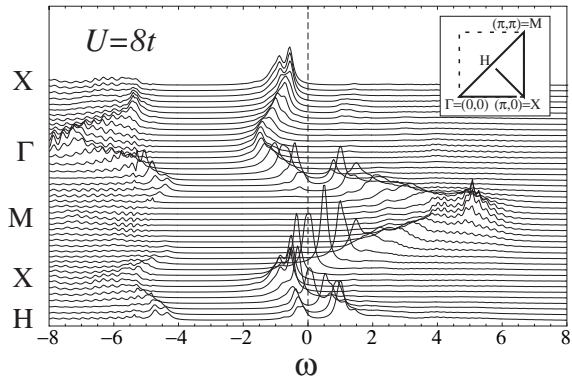}
\caption{Single-particle spectral weight obtained by CPT at $n=1.125$ as a
function of energy $\protect\omega$ in units of $t$ for wave vectors along
the high symmetry directions shown in the inset. Band parameters are $%
t^{\prime}=-0.3t$, $t^{\prime\prime }=0.2t$.}
\label{dispersion}
\end{figure}

The physical mechanism for the pseudogap here is that those quasiparticles
in regions of the Fermi surface that are connected to other such regions by
wave vectors that have a broad spread of radius $\delta $ around $(\pi ,\pi )
$ suffer strong scattering by short range correlations~\cite{st2004}. Thus,
the pseudogap driven by the Mott physics with short range correlations
occurs around zero energy and only in these regions of the Fermi surface
(hot spots). In contrast, the Mott gap occurs for all wave vectors and is
not tied to zero energy. This is illustrated on Fig.~\ref{dispersion}, which
shows energy dispersion curves obtained by CPT at $n=1.125$ for wave vectors
along the high symmetry directions shown in the inset. The range of
frequencies away from zero energy where $A(\mathbf{k},\omega )=0$ for all
wave vectors is the Mott gap. At finite electron doping it always opens up
at negative energies when the Hubbard coupling $U$ is sufficiently large,
and the chemical potential lies in the upper Hubbard band. The range of
frequencies around zero energy where spectral weight is suppressed only for
some wave vectors along $\Gamma $-M and X-H directions is the pseudogap.
This suppression of spectral weight at zero energy is also seen in the Fermi
surface plot for the same parameters [Fig.~\ref{FS_plot}(c)]. Note that the
longer range AFM correlations, which are observed in experiment~\cite%
{Mang2003}, would probably only reinforce the strong-coupling mechanism that
already exists in the presence of short-range correlations. One may
speculate that at strong coupling the condition $\xi >\xi _{th}$ becomes $%
\xi >a$ (with $a$ the lattice spacing), which is easier to satisfy. This
would connect the weak and strong coupling regimes in a continuous manner.

We acknowledge useful discussions with M.~Greven. The present work was
supported by NSERC (Canada), FQRNT (Qu\'{e}bec), CIAR, RQCHP, and the Tier I
Canada Research Chair Program (A.-M.S.T.).

\end{document}